\documentclass[twocolumn,showpacs,preprintnumbers,amsmath,amssymb]{revtex4}

\usepackage{psfig}
\usepackage{graphicx}
\usepackage{dcolumn}
\usepackage{bm}
\usepackage[dvips]{color}

\newcommand{\be}{\begin{equation}}
\newcommand{\ee}{\end{equation}}

\begin{document}
\title{Tailored quantum dots for entangled photon pair creation}

\author{A. Greilich}
\author{M. Schwab}
\author{T. Berstermann}
\author{T. Auer}
\author{R. Oulton}
\author{D.~R. Yakovlev}
\author{M. Bayer}
\affiliation{Experimentelle Physik II,
             Universit\"at Dortmund,
             D-44221 Dortmund, Germany}

\author{V. Stavarache}
\author{D. Reuter}
\author{A. Wieck}
\affiliation{Angewandte Festk\"orperphysik,
             Ruhr-Universit\"at Bochum,
             D-44780 Bochum, Germany}

\date{\today}

\begin{abstract}
We compare the asymmetry-induced exchange splitting $\delta_1$ of
the bright-exciton ground-state doublet in self-assembled
(In,Ga)As/GaAs quantum dots, determined by Faraday rotation, with
its homogeneous linewidth $\gamma$, obtained from the radiative
decay in time-resolved photoluminescence. Post-growth thermal
annealing of the dot structures leads to a considerable increase
of the homogeneous linewidth, while a strong reduction of the
exchange splitting is simultaneously observed. The annealing can
be tailored such that $\delta_1$ and $\gamma$ become comparable,
whereupon the carriers are still well confined. This opens the
possibility to observe polarization entangled photon pairs through
the biexciton decay cascade.
\end{abstract}


\maketitle

Entangled photon pairs are a key requirement for the
implementation of quantum teleportation schemes. \cite{Zeilinger}
Typically, such photon pairs are created by parametric down
conversion of a strongly attenuated laser beam in a non-linear
optical crystal, with limited efficiency. Recently, the decay of a
biexciton complex confined in a quantum dot (QD) has been
suggested as an efficient source for polarization entangled photon
pairs. \cite{BensonPRL00} This concept was based on the assumption
of an idealistic QD structure for which the valence band ground
state has pure heavy hole character with angular momentum
projections $J_{h,z} = \pm 3/2$ along the heterostructure growth
direction. When an electron-hole pair is injected, the momenta of
the carriers become coupled by the exchange interaction. If the
dot has perfect $D_{2d}$-symmetry, angular momentum is a good
quantum number: the optically active states with momenta $M =
\pm$1 are degenerate, and their decay leads to emission of
$\sigma^{\pm}$-circularly polarized photons.

\begin{figure}[tbp]
\centering
\centerline{\psfig{figure=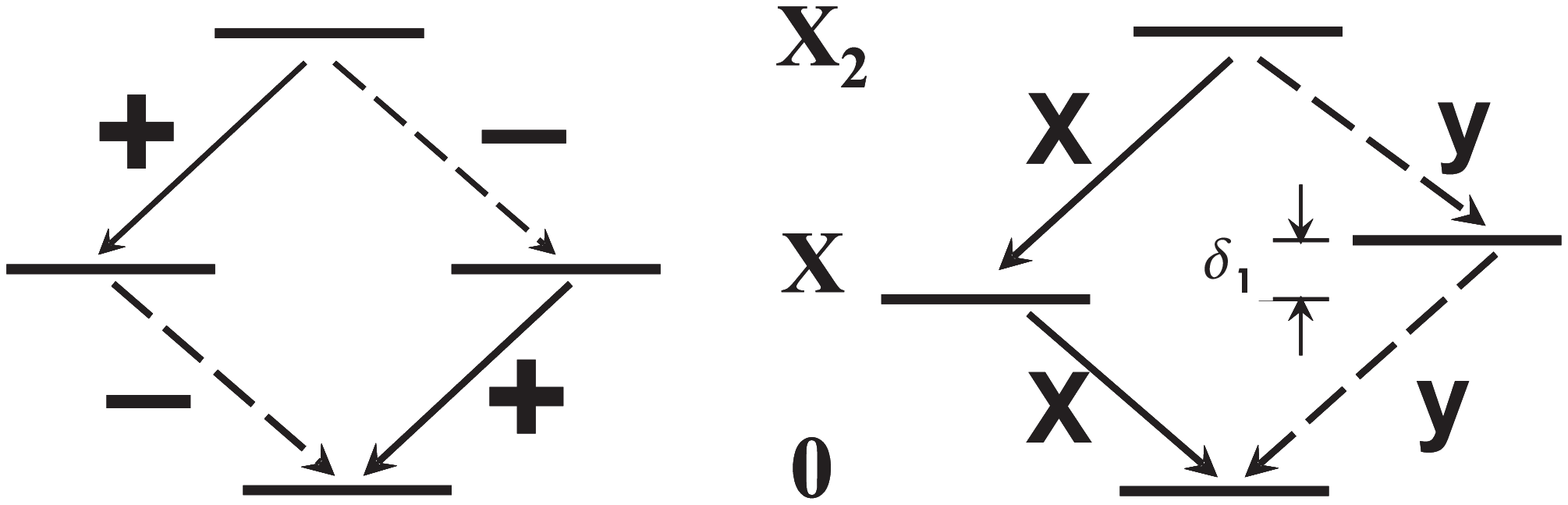,width=8.5truecm}}
\centerline{\psfig{figure=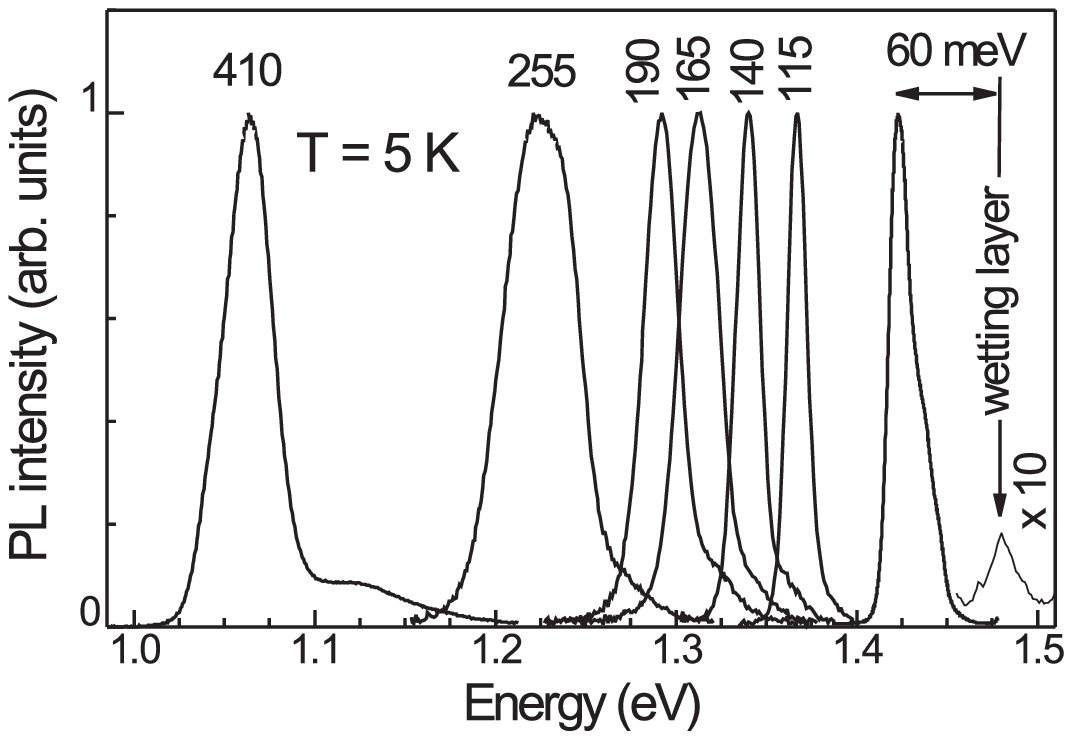,width=8.5truecm}}
\caption{Upper: Scheme of the possible decay channels of a
biexciton $X_2$ confined in a QD. Left hand side gives the
situation for an idealized dot with $D_{2d}$-symmetry, while the
right hand side does the same for a dot with reduced symmetry.
Plus/minus signs and x/y signs indicate circular and linear
polarization of the emitted photons, respectively. Lower: Low
excitation photoluminescence spectra of QD ensembles annealed at
different temperatures, leading to different confinement
potentials, as indicated at each spectrum.} \label{fig1}
\end{figure}

If the dot ground states are occupied by two electrons and two
holes, each with opposite spin orientations, a spin singlet
biexciton $X_2$ is formed, for whose decay two channels exist, as
shown in Fig.~1 (upper panel left). The first photon is emitted with
either $\sigma^+$ or $\sigma^-$-polarization, and then the second photon
with opposite polarization, as long as no spin flip occurs after
the first process. Unless a polarization measurement is
performed, the two photon polarization state is therefore
described by $ \mid 2 \gamma \rangle = \left( \mid + \rangle_1
\mid - \rangle_2 + \mid - \rangle_1 \mid + \rangle_2 \right) /
\sqrt{2}$, forming an entangled state. A key requirement is that
the photons emitted at each stage of the cascade are
quasi-degenerate within their homogeneous linewidth, such that
they cannot be distinguished by an energy measurement.

Experiments have failed up to now to demonstrate such an
entanglement, as only classical correlations were observed.
\cite{noentanglement} While some of the idealizations of the
original proposal are well fulfilled, for example, for strongly
confined self-assembled (In,Ga)As/GaAs quantum dots (such as the
long exciton spin relaxation time as compared to the radiative
lifetime \cite{PaillardPRL01}, or the almost pure heavy-hole
character of the valence band ground state \cite{CortezPRB01}), a
fundamental problem arises from the broken $D_{2d}$ symmetry,
which is reduced to at least $C_{2v}$ or even lower symmetry in
realistic dot structures. \cite{BayerPRL99,BayerPRB02,BesterPRB05}
As a consequence, angular momentum is no longer a good quantum
number and the $\pm$ 1 excitons become mixed to linearly polarized
eigenstates, resulting in an energy splitting $\delta_1$ of the
bright exciton doublet (see Fig. 1, upper panel right). Generally,
this splitting is considerably larger than the homogeneous
linewidth of the exciton. Therefore the two decay channels can be
distinguished even without a polarization measurement, simply by
measuring the energy of the first photon. Photon entanglement is
not preserved in this case.

The only way to achieve polarization entanglement is to reduce the
splitting such that $\delta_1$ becomes smaller than the
homogeneous linewidth $\gamma$, which at cryogenic temperatures is
radiatively limited. \cite{LangbeinPRB04a} Several strategies have
been pursued to reach the goal of a quasi-degeneracy of the bright
exciton doublet. One example is the application of an electric
field in the quantum dot plane, to compensate the asymmetry. This
has shown some promise, but the reduction of $\delta_1$ was still
too small. \cite{KowalikAPL05} Lately, it was found by non-linear
optical techniques that thermal annealing may lead to a strong
reduction of the exchange splitting down to the few $\mu$eV-range.
\cite{LangbeinPRB04b,TartakovskiiPRB04} Very recently, it has been
demonstrated by single dot spectroscopy that, within the
experimental accuracy, the splitting may even become zero or its
sign may be reversed. \cite{YoungPRB05}

Here we complete this picture by addressing not only the asymmetry
splitting for dots with varying confinement, but we also compare
this splitting to the homogeneous linewidth $\gamma$: Any
reduction of $\delta_1$ even to very small values would not enable
entanglement as long as $\delta_1 > \gamma$. We show that the two
energies may be made comparable through an annealing step, which,
however, still keeps the dot carriers well confined. For
determining the energies we use spectroscopic techniques
complementary to those used previously, and we compare the results
to data reported in literature.

The experiments were performed on arrays of self-assembled
(In,Ga)As/GaAs QDs. To obtain strong enough light-matter
interaction, the samples contained 20 layers of QDs, that were
separated by 60 nm wide barriers. The structures were fabricated
by molecular beam epitaxy on (001)-oriented GaAs substrate. The
samples were annealed for 30 s at different temperatures $T_{ann}$
between 800 and 980$^{\circ}$C by which the confinement is reduced
due to intermixing between dot and barrier material. Fig. 1 (lower
panel) shows typical photoluminescence spectra of samples
differing in $T_{ann}$, which show the established behavior for
such a series of structures: \cite{annealing} With increasing
$T_{ann}$, a blue shift as well as a narrowing of the emission
line is observed. Even though there is a clear trend of increasing
emission energy with increasing $T_{ann}$ for the samples under
study, an exact correlation cannot be made. As a more
characteristic quantity for the electronic confinement therefore
we in the following label the samples by the confinement
potential, which we define as energy separation of the ground
state dot emission from the wetting layer at about 1.48 eV. The
wetting layer emission also shifts to higher energies with increasing
$T_{ann}$, but this shift is weak. The confinement energies range
from 410 down to 60 meV.

\begin{figure}[tbp]
\centering
\centerline{\psfig{figure=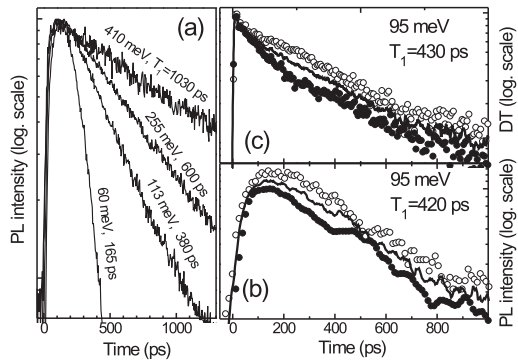,width=\columnwidth}}
\caption{(a) Time-resolved photoluminescence of QD ensembles with
different confinement without polarization analysis. Excitation
power was 1 W cm$^{-2}$.
(b) Same as (a), but for a
sample with 95 meV confinement. Excitation into wetting layer was
$\sigma^+$-polarized, detection was either $\sigma^+$
(open circles) or $\sigma^-$-polarized (solid circles). Solid line
gives average of both traces.
(c) Differential transmission
for same sample as in panel (b). Pump beam was
$\sigma^+$-polarized, probe either $\sigma^+$ (open circles) or
$\sigma^-$-polarized (solid circles). Solid line again gives
average.} \label{fig2}
\end{figure}

The samples were immersed in helium gas at a temperature of 5 K.
Optical excitation was done by a mode-locked Ti-sapphire laser
emitting pulses with a duration of about 1 ps at 75.6 MHz
repetition rate, which hit the sample along the heterostructure
growth direction.

The exciton lifetimes were studied using time-resolved
photoluminescence, for which the wavelength of the pulsed laser
was tuned to the GaAs band gap at 1.512 eV. The emission was
dispersed by a 0.5 m monochromator and detected by a streak camera
with a S1 photocathode. Excitation powers as low as possible were
used to study the pure exciton decay by avoiding multiparticle
occupation effects such as Pauli blocking and related ground shell
refilling. Fig.~2(a) shows decay curves of four different QD
samples. The excitation was
linearly polarized, while the emission was detected without
polarization resolution. One clearly sees that the decay is the
faster, the shallower the confinement is. The observed decays have
been analyzed by single exponential fit, and the decay times
are indicated at each trace. The decay time decreases from 1030 to
165 ps with decreasing confinement. This behavior is expected: the
exciton lifetime is determined by the exciton coherence volume,
which is given by the dot size. The size is increased by the
annealing step, leading to the decrease of the exciton lifetime.

Fig. 2(b) shows time-resolved PL spectra of a sample with a
confinement potential of 95 meV. In this case the excitation was
$\sigma^+$-circularly polarized and resonant with the wetting layer at about 1.48
eV. Detection was taken either $\sigma^+$ (open circles) or
$\sigma^-$ (solid circles) polarized. For clarity, the two curves
have been shifted vertically relative to each other. The average
of these traces is also shown (the solid line) which follows to a
good approximation an exponential dependence with a decay time of
420 ps. The polarization-resolved decays show some modulation
which is in antiphase for the two curves. A rough estimate gives a
period of 600$\pm$50 ps for the oscillation, which corresponds to
an energy splitting of about 7 $\mu$eV. This is comparable to the
expected fine structure splitting $\delta_1$.
\cite{LangbeinPRB04b}

A priori it is not clear that time-resolved PL measurements yield
the exciton lifetime $T_1$, in particular for non-resonant
excitation, as the dynamics involves also carrier relaxation.
Therefore we have also performed pump-probe differential transmission studies
on the 95 meV confinement sample: The pump beam was resonant to
the ground state dots transition and excites a corresponding carrier
population, whose decay is then tested by a probe beam. The data
are shown in Fig.~2(c). The traces show data with $\sigma^+$ pump
excitation, and $\sigma^+$ (open circles) and $\sigma^-$ (solid circles)
-polarized detection, respectively. The solid line gives again the
average of the two traces, resulting in a decay time of 430 ps.
Within the experimental error, this time
agrees well with that obtained by time-resolved photoluminescence,
confirming that under the applied experimental conditions, the PL
decay time does indeed give the exciton lifetime $T_1$, from which
the homogeneous linewidth $\gamma = 2 \hbar / T_1$ is obtained
(see Fig.~4). Note that for the polarized differential transmission traces
an antiphase modulation is again observed, even though it is too
weak to determine an oscillation period.

\begin{figure}[tbp]
\centering
\centerline{\psfig{figure=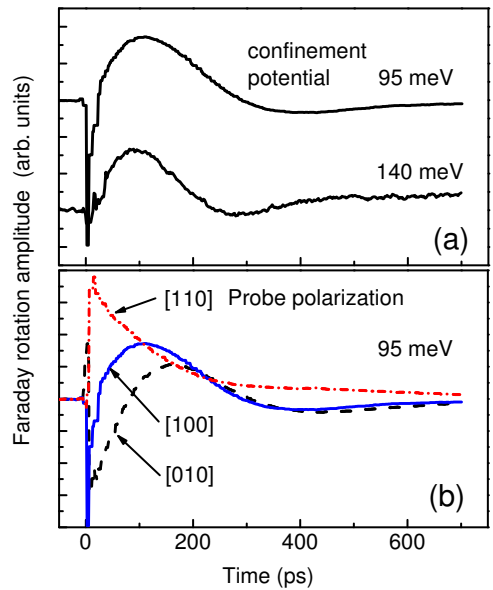,width=7truecm}}
\caption{(a) Faraday rotation signal with probe beam polarized
linearly along the [100] crystal direction for samples with confinement
potentials of 95 and 140 meV. (b) Faraday
rotation traces recorded on the sample with 95 meV confinement
potential for different linear polarization orientations of probe
beam. In both cases, the pump beam was $\sigma^+$-polarized. $T$ =
5 K.} \label{fig3}
\end{figure}

As mentioned, studies of the exchange interaction induced
splitting have been up to now reported by photoluminescence on single
QDs \cite{BayerPRL99,BayerPRB02} or by non-linear spectroscopy
such as four-wave-mixing \cite{LangbeinPRB04b} or differential
transmission \cite{TartakovskiiPRB04} on ensembles. Here we use
another non-linear technique to address this problem, namely
pump-probe Faraday rotation \cite{Awschalom}, for which the laser
was tuned to the energy of the ground state transition in the QDs
and split into two trains: An electron spin polarization is
induced by a circularly polarized pump beam and is tested by the
rotation of the linear polarization of a probe beam. For recording
the rotation angle, a homodyne technique based on phase-sensitive
balanced detection was used.

Following previous results for small $\delta_1$ values
\cite{LangbeinPRB04b}, we focused in these experiments on
strongly annealed dots with small confinement potentials.
Fig.~3(a) shows Faraday rotation signals for two samples with
confinement potentials of 140 and 95 meV.
The polarization direction of the probe beam was directed
along the [100] crystal axis. Strongly damped oscillations are
observed in both cases (note that the data were recorded at zero
magnetic field), as a result of precession of the exciton angular
momentum about the in-plane anisotropy axes of the QDs along [110]
and [1$\bar{1}$0]. This precession reflects the quantum beats
occurring due to the coherent excitation of both linearly
polarized exciton eigenstates $\mid \Psi_{1, 2} \rangle \propto
\mid + 1 \rangle \pm \mid -1 \rangle$ by the circularly polarized
pump pulse. The data can be analyzed by an oscillatory fit
function with an exponentially damped amplitude. The oscillation
period is clearly shorter for the sample with a stronger
confinement potential (390 ps) as compared to the sample with
weaker confinement (600 ps). The energy splittings $\delta_1$ that
are derived from the oscillation period are 10.5 and 7.1 $\mu$eV
for the 140 and 95 meV samples, respectively.
These values agree very well with data determined
from four-wave-mixing on samples with comparable confinement.
\cite{LangbeinPRB04b}

For clarity we note that in the pump-probe measurements true
quantum beats are observed due to direct resonant excitation of
the exchange-split ground state exciton, while in the
time-resolved photoluminescence studies with non-resonant
excitation described above the observed oscillation are due to
polarization interferences of the emitted photons.
\cite{SenesPRB05}

For understanding the observed Faraday rotation more intuitively,
a pseudospin formalism can be used for the exciton description.
\cite{IVCHENKObook} The bright exciton doublet with $\mid M \mid$
= 1 is described by the matrix
\begin{eqnarray*}
{\cal H} = \left(
       \begin{array}{cc}
       E_0 + \frac{\delta_0}{2}  &  \frac{\delta_1}{2}   \\
       \frac{\delta_1}{2}   &  E_0 + \frac{\delta_0}{2}  \\
       \end{array}
\right),
\end{eqnarray*}
where $E_0$ is the exciton energy, disregarding exchange
interaction effects, and $\delta_0$ gives the energy splitting
between the bright excitons with $\mid M \mid = 1$ and the
dark excitons with $\mid M \mid = 2$. This Hamiltonian can be
rewritten as ${\cal H} = \left( E_0 + \delta_0/2 \right) I +
\delta_1 \sigma_x$/2, where $I$ is the identity and $\sigma_x$ is
the Pauli matrix. The second term has the same form as that of a
spin in a perpendicular magnetic field pointing along the
$x$-direction, resulting in the observed precession.

Fig. 3(b) shows different Faraday rotation traces of the
QD sample with 95 meV confinement. The polarization direction of
the probe beam, as indicated by the corresponding crystal
directions, was varied. The probe polarization has no influence on
the beat period, but evidently affects the phase of the
oscillation. The exciton state created by the circularly polarized
light is given by the superposition state
\begin{eqnarray}
\Psi \propto \Psi_1 \exp \left( - i \frac{\delta_1 t}{2 \hbar}
\right) + \Psi_2 \exp \left( + i \frac{\delta_1 t}{2 \hbar}
\right).
\end{eqnarray}
At a certain time, different components of this state are tested
by varying the polarization of the probe, causing the phase shift
of the Faraday rotation. In particular, a $\pi-$phase shift occurs
for a $90^0$-deg. rotation of the probe polarization from [010] to [110].

\begin{figure}[tbp]
\centerline{\psfig{figure=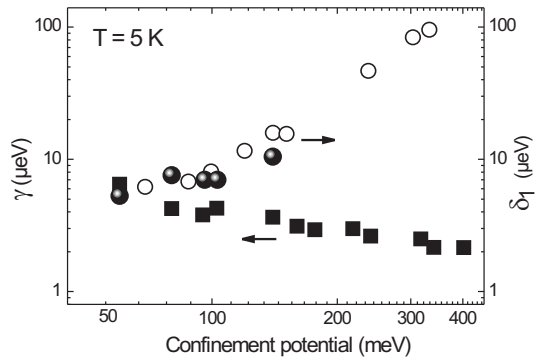,width=8truecm}}
\caption{Homogeneous linewidth $\gamma$ (squares) and exchange
splitting $\delta_1$ (circles) of the exciton in
InGaAs/GaAs quantum dots as function of the confinement potential.
Solid circles are measured with Faraday rotation; open
circles are taken from Ref.~\cite{LangbeinPRB04b}. Squares are
measured by time resolved PL.} \label{fig4}
\end{figure}

Fig.~4 gives an overview of $\gamma$ and
$\delta_1$ as function of the confinement
potential. Note that both values are plotted on a logarithmic
scale, as is the confinement potential. The solid circles and solid
squares give the exchange splitting and homogeneous linewidth, respectively,
measured in our experiments. For comparison,
the open circles give the $\delta_1$ values from Ref.
\cite{LangbeinPRB04b}. For the as-grown QD sample with the highest
confinement, $\delta_1$ is by a factor 50 larger than the homogeneous
linewidth. This holds for all samples with confinement potentials
larger than about 300 meV. Most studies reported up to now on
polarization entangled photons were done on comparable structures.
The data in Fig.~4 underlines why no entanglement was observed yet.

However, when the confinement potential height is decreased
to less than 300 meV, a drastic reduction of the exchange splitting is observed.
Simultaneously, the homogeneous linewidth
increases. For the QDs with confinement potential around
100 meV, both quantities are of comparable magnitude. For the most
shallow QDs, $\delta_1$ is reduced down to about 5 $\mu$eV, while
$\gamma$ is increased to 7 $\mu$eV. For these QDs the exchange
splitting can no longer be resolved by an energy measurement. It
is these QDs from which polarization entangled photon pairs might
be expected, and which we will study in the future. For this
purpose the cross-correlation should be measured for the biexciton
decay cascade in a single QD by a Hanbury-Brown-Twiss setup.
\cite{noentanglement}

In summary, we have demonstrated that thermal annealing performed
on QDs favors a situation in which polarization entangled photon
pairs may be observed. On one hand, it reduces strongly the
asymmetry-induced exchange splitting of excitons. On the other
hand, it simultaneously increases the exciton homogeneous
linewidth, under which a finite splitting may be hidden. The
latter effect may be enhanced further if the QDs were placed in an
optical resonator, which could be used not only for reducing the
exciton lifetime through the Purcell effect \cite{PurcellPR46},
but also for funnelling the emission into a desired spatial
direction, thus enhancing the collection efficiency.
\cite{PurcellPR46,GerardPRL98,OhnesorgePRB97}

{\bf Acknowledgements.} We are grateful to E.L. Ivchenko for
insightful discussions. The work was supported by the DFG
(research group 'Quantum Optics of Semiconductor
Heterostructures'). R. O. thanks the Alexander von Humboldt foundation.


\begin{thebibliography}{99}

\bibitem{Zeilinger} D. Bouwmeester, A. Ekert, and A. Zeilinger, {\sl The
Physics of Quantum Information}, Springer, Berlin (2000).

\bibitem{BensonPRL00} O.~Benson, C.~Santori, M.~Pelton, Y.~Yamamoto, Phys. Rev. Lett. {\bf 84}, 2513 (2000).

\bibitem{noentanglement}
see, for example, E.~Moreau, I.~Robert, L.~Manin, V.~Thierry-Mieg, J.~M.~Gerard, I.~Abram, Phys. Rev. Lett. {\bf
87}, 183601 (2001); C.~Santori, D.~Fattal, M.~Pelton, G.~S.~Solomon, Y.~Yamamoto, Phys. Rev. B {\bf 66}, 45308 (2002); R.~M.~Stevenson, R.~M.~Thompson, A.~J.~Shields, I.~Farrer, B.~E.~Kardynal, D.~A.~Ritchie, M.~Pepper, Phys. Rev. B {\bf 66}, 081302(R) (2002).

\bibitem{PaillardPRL01}
M.~Paillard, X.~Marie, P.~Renucci, T.~Amand, A.~Jbeli, and J.~M.~Gérard,
Phys. Rev. Lett. {\bf 86}, 1634 (2001).

\bibitem{annealing}
see, for example, S. Fafard and C. Allen, Appl. Phys. Lett. {\bf
75}, 2374 (1999).

\bibitem{CortezPRB01}
see, for example, S. Cortez, O. Krebs, P. Voisin, and J.~M.
Gerard, Phys. Rev. B {\bf 63}, 233306 (2001).

\bibitem{BayerPRL99}
M.~Bayer, A.~Kuther, A.~Forchel, A.~Gorbunov, V.~B.~Timofeev, F.~Schäfer, J.~P.~Reithmaier,
T.~L.~Reinecke and S.~N.~Walck, Phys. Rev. Lett. {\bf 82}, 1748 (1999).

\bibitem{BayerPRB02}
see, for example, M.~Bayer, G.~Ortner, O.~Stern, A.~Kuther, A.~A.~Gorbunov, A.~Forchel,
P.~Hawrylak, S.~Fafard, K.~Hinzer, T.~L.~Reinecke, S.~N.~Walck,
J.~P.~Reithmaier, F.~Klopf, and F.~Schäfer, Phys. Rev. B {\bf 65},
195315 (2002) and references therein.

\bibitem{BesterPRB05}
see, for example, G. Bester and A. Zunger, Phys. Rev. B {\bf 71},
045318 (2005).

\bibitem{LangbeinPRB04a}
W.~Langbein, P.~Borri, U.~Woggon,
V.~Stavarache, D.~Reuter, and A.~D.~Wieck,
Phys. Rev. B {\bf 70}, 033301 (2004).

\bibitem{KowalikAPL05}
K.~Kowalik, O.~Krebs, A.~Lemaître, S.~Laurent, P.~Senellart, P.~Voisin, and J.~A.~Gaj,
Appl. Phys. Lett. {\bf 86}, 041907 (2005).

\bibitem{LangbeinPRB04b}
W.~Langbein, P.~Borri, U.~Woggon,
V.~Stavarache, D.~Reuter, A.~D.~Wieck, Phys. Rev. B {\bf 69}, 161301(R) (2004).

\bibitem{TartakovskiiPRB04}
A.~I.~Tartakovskii, M.~N.~Makhonin, I.~R.~Sellers, J.~Cahill, A.~D.~Andreev, D.~M.~Whittaker,
J-P.~R.~Wells, A.~M.~Fox, D.~J.~Mowbray, M.~S.~Skolnick, K.~M.~Groom, M.~J.~Steer, H.~Y.~Liu, and M. Hopkinson,
Phys. Rev. B {\bf 70}, 193303 (2004).

\bibitem{YoungPRB05}
R.~J.~Young, R.~M.~Stevenson, A.~J.~Shields, P.~Atkinson, K.~Cooper, D.~A.~Ritchie,
K.~M.~Groom, A.~I.~Tartakovskii, and M.~S.~Skolnick, Phys. Rev. B {\bf 72}, 113305 (2005).

\bibitem{Awschalom} D.~D. Awschalom and N. Samarth, in {\sl
Semiconductor Spintronics and Quantum Computation}, edited by
D.~D. Awschalom, D. Loss, and N. Samarth, Springer Berlin (2002).

\bibitem{SenesPRB05}
For a detailed discussion see, for example,
M.~Sénès, B.~Urbaszek, X.~Marie, T.~Amand,
J.~Tribollet, F.~Bernardot, C.~Testelin, M.~Chamarro, and
J.-M.~Gérard, Phys. Rev. B {\bf 71}, 115334 (2005), and references therein.

\bibitem{IVCHENKObook} E.L. Ivchenko, {\sl Optical Spectroscopy of
semiconductor nanostructures}, Alpha Science International, Harrow
(2005).

\bibitem{PurcellPR46} E.M. Purcell, Phys. Rev. {\bf 69}, 681 (1946).

\bibitem{GerardPRL98}
J.~M.~Gérard, B.~Sermage, B.~Gayral, B.~Legrand, E.~Costard, and V.~Thierry-Mieg,
Phys. Rev. Lett. {\bf 81}, 1110 (1998).

\bibitem{OhnesorgePRB97} B. Ohnesorge, M. Bayer, A. Forchel, J. P. Reithmaier, N. A. Gippius, and S. G. Tikhodeev, Phys. Rev. B {\bf 56}, R4367 (1997).

\end{thebibliography}
\end{document}